\documentclass[runningheads]{llncs}
\ifdefined\pdfsuppressptexinfo\pdfsuppressptexinfo=-1\fi
\usepackage[T1]{fontenc}
\usepackage{amsmath}
\usepackage{newtxtext}
\usepackage[varvw]{newtxmath}
\usepackage{graphicx}
\usepackage{booktabs}
\usepackage[hidelinks]{hyperref}
\usepackage{color}

\graphicspath{{figures/}}
\begin{document}

\title{Data-Driven Prediction of Optimal MPI Rank Placement and Node Count for Memory-Bound CFD on AMD EPYC Clusters}
\titlerunning{Data-Driven Rank Placement and Node Count for CFD on AMD EPYC}

\author{Marcin Lawenda\inst{1}\orcidID{0000-0003-4844-3655} \and
Łukasz Szustak\inst{2}\orcidID{0000-0001-7429-6981} \and
Aleksandra Krasicka\inst{1}\orcidID{0009-0008-3860-4439}}
\authorrunning{M. Lawenda et al.}

\institute{Poznan Supercomputing and Networking Center, Jana Pawła II 10, 61-139 Poznań, Poland\\
\email{\{lawenda,akrasicka\}@man.poznan.pl} \and
Czestochowa University of Technology, Dąbrowskiego 69, 42-201 Częstochowa, Poland\\
\email{lukasz.szustak@pcz.pl}}

\maketitle

\begin{abstract}
Users who run a memory-bound OpenFOAM case on an AMD EPYC cluster must decide how many nodes to use and how many MPI ranks to place on each core complex. The fastest choice is often not the cheapest one. This paper revisits our earlier analytical model of this performance and energy trade-off with 560 measured configurations of the motorBike case on five Zen~4 and Zen~5 EPYC processors, including socket power from processor telemetry. On up to four nodes, small meshes are both fastest and cheapest on a single node on most processors. For the largest mesh the largest node count is both the fastest and the cheapest option, because its speedup is superlinear. A genuine trade-off arises only for medium meshes, where full speed costs a median of 20 to 97\% more node hours than the cheapest option. The L3 miss rate and the halo traffic per cell are consistent with this pattern. For a new mesh of the same case, the configuration that was optimal on the nearest measured mesh is near optimal in 75 to 85\% of the decisions. Running the two configurations it recommends for speed and for cost raises this share to 85 to 90\% for runtime and node hours and determines the trade-off with median errors of 3 to 5 percentage points. Neither gradient boosted trees trained on the same data nor the earlier polynomial model improves on this simple procedure.
\end{abstract}

\section{Introduction}

Computational fluid dynamics (CFD) codes such as OpenFOAM are limited by memory traffic more than by arithmetic. On current AMD EPYC processors their speed depends on how well the cells of each MPI rank fit into the L3 cache of a core complex (CCX) and on how much halo data the ranks exchange. Before submitting such a job, a user has to decide how many compute nodes to request and how many ranks to place on each CCX. Both choices affect the runtime as well as the cost of the job in node hours or in energy, so the cost of a shorter runtime should be known before the job starts.

Our earlier work studied this question on a cluster of AMD EPYC Milan processors \cite{law2025prediction}. It introduced the CVOPTS metric, the number of cells processed per core and time step, fitted it with a polynomial in the number of cells per core and derived the runtime and the cost of each node count from the fit. Since that cluster did not measure power, the energy was estimated from the thermal design power (TDP). A later study compared several EPYC generations on CFD applications and traced their differences to the cache and memory subsystems \cite{law2025evaluating}.

This paper extends that work to five processors of the Zen~4 and Zen~5 generations, uses socket power from processor telemetry and treats the rank placement as a second decision. It addresses a user with a new mesh of a case that has already been measured. The paper makes the following four contributions to this question:
\begin{itemize}
\item Measurements of the speed and cost trade-off over eight meshes, up to four nodes and five placements per processor, which show for which meshes a trade-off exists and how large it is.
\item An analysis of hardware counters that relates superlinear speedup to the L3 miss rate and the loss of efficiency to the halo traffic per cell.
\item A simple prediction procedure for a new mesh, which takes the configuration from the nearest measured mesh and measures the trade-off with two guided runs.
\item An assessment of this procedure against gradient boosted trees, the earlier polynomial model, the best of random runs and common fixed rules.
\end{itemize}
Section~\ref{sec:related} reviews related work, Section~\ref{sec:setup} describes the measurements and Section~\ref{sec:formulation} defines the trade-off. Sections~\ref{sec:measured} and \ref{sec:regimes} present the measured trade-off and its hardware causes. Section~\ref{sec:method} introduces the prediction method, Section~\ref{sec:predictions} shows its predictions and Section~\ref{sec:assessment} assesses them. Section~\ref{sec:conclusions} concludes.

\section{Related Work}
\label{sec:related}

Choosing the node count and the placement of a CFD job relates to three areas of research. Work on memory-bound CFD codes explains their performance on multicore nodes. Work on process placement and resource choice optimises the mapping of a job for performance and energy. Learned performance models and autotuning predict or search for good configurations from measurements. The following subsections review these areas in turn and position the present paper within each.

\subsection{Memory-Bound CFD Performance}
Implicit CFD codes on unstructured meshes have long been known to be limited by memory bandwidth more than by floating point throughput. Gropp et al. analysed a PETSc based Euler solver from this perspective and showed that its sequential and parallel performance follows from the data volume moved per cell \cite{gropp2000fun3d}. Models of a single node make such limits quantitative. The roofline model bounds the performance by the arithmetic intensity and the memory bandwidth \cite{williams2009roofline}. The Execution-Cache-Memory model adds the transfers between the cache levels and thereby predicts the gain when the data of a process fits into a cache \cite{stengel2015ecm}. When the working set per process drops into the cache as more processes share a fixed problem, the speedup can even exceed the number of processes \cite{ristov2016superlinear}. Scaling studies on large systems analysed the strong and weak scaling of OpenFOAM together with the cost of its MPI communication \cite{axtmann2016openfoam}. CFD microbenchmarks have been compared across diverse HPC architectures \cite{galeazzo2024cfd}. Our earlier study related the performance of EPYC processors on CFD codes to the size and the organisation of the L3 cache \cite{law2025evaluating}. Our CVOPTS model predicted the cost of each node count, for one processor type with one placement and a fixed power per socket \cite{law2025prediction}. The present paper measures both effects on the same meshes and uses them to explain where a trade-off between speed and cost appears.

\subsection{Process Placement and Resource Choice}
On hierarchical nodes the placement of processes determines which caches and memory controllers they share. Libraries such as hwloc expose this topology to runtime systems \cite{broquedis2010hwloc}. Mapping algorithms place communicating processes close to each other inside a node \cite{jeannot2010treematch} or across the network \cite{hoefler2011mapping}, with the number of processes fixed. In this paper the number of processes per core complex and the number of nodes are the decisions themselves. Choosing the concurrency of an application for power and performance has been studied for multithreaded codes, where prediction models select the number of threads, their placement and the processor frequency together \cite{curtismaury2008prediction}. For MPI codes the waiting time of processes off the critical path can be used to lower the frequency with little loss of time \cite{rountree2009adagio}, while power caps trade performance against energy at run time \cite{krzywaniak2022depo}. At the level of the whole system, energy has been addressed by scheduling, resource allocation and power management \cite{czarnul2019survey}, where the power of a job can be predicted from historical data to guide the scheduler \cite{borghesi2016predictive}. These approaches act on the side of the system or of the runtime. The present work supports the user, who chooses the nodes and the placement of one job before it starts.

\subsection{Learned Models and Autotuning}
Regression models learned from measurements predict the performance of a configuration without running it \cite{lee2006accurate,ipek2006efficiently}. Gradient boosted trees \cite{chen2016xgboost} are a common choice for such tabular data. Several machine learning methods have been benchmarked for the performance modelling of scientific applications \cite{malakar2018benchmarking}. Automated empirical modelling fits scaling laws in the number of processes or the problem size, as done by Extra-P to find scalability bugs \cite{calotoiu2013extrap}. Transfer learning reuses inexpensive measurements, such as runs at small scale, to model an expensive setting with few runs \cite{marathe2017transfer}. Autotuning frameworks search a configuration space by repeated runs \cite{ansel2014opentuner}, with Bayesian optimisation and multitask learning across related tasks \cite{liu2021gptune} or with energy as a target \cite{wu2025ytopt}. Profiling of the hardware and of the communication also guides the choice between task and data parallelism for applications outside CFD \cite{szu2023pyprofiling}. These methods need either a model trained on many configurations or a series of runs of the new case. The present paper examines to what extent the measurements of a few meshes of a case transfer to a new mesh. It compares a direct transfer of the optimal configuration with a learned surrogate.

\section{Experimental Setup}
\label{sec:setup}

\subsection{Hardware}
The experiments use five AMD EPYC processor models of two generations (Table~\ref{tab:cpus}). Each node holds two processors of the same model. Four nodes are available for each model except Turin-128, which offers two. The Genoa-X and Turin-96 nodes are connected by InfiniBand at 200~Gb/s, while the other platforms use 400~Gb/s. All nodes run Rocky Linux~9 with firmware and power settings chosen by the provider. Turin-c is the dense Zen~5c model with 16 cores per CCX, twice as many as in the other models. Genoa-X adds stacked cache dies that triple the L3 cache of Genoa. The memory speed listed in Table~\ref{tab:cpus} is that of the installed modules.

\begin{table}
\caption{Processors used in this study (one socket).}
\label{tab:cpus}
\centering
\setlength{\tabcolsep}{4pt}
\begin{tabular}{lccccc}
\toprule
 & Genoa & Genoa-X & Turin-96 & Turin-128 & Turin-c \\
\midrule
Model & 9654 & 9684X & 9655 & 9755 & 9965 \\
Microarchitecture & Zen 4 & Zen 4 & Zen 5 & Zen 5 & Zen 5c \\
Cores & 96 & 96 & 96 & 128 & 192 \\
Cores per CCX & 8 & 8 & 8 & 8 & 16 \\
L3 cache (MB) & 384 & 1152 & 384 & 512 & 384 \\
L3 per core (MB) & 4 & 12 & 4 & 4 & 2 \\
Memory (DDR5 MT/s) & 4800 & 4800 & 6000 & 5600 & 6400 \\
Peak memory bandwidth (GB/s) & 461 & 461 & 576 & 538 & 614 \\
TDP (W) & 360 & 400 & 400 & 500 & 500 \\
\bottomrule
\end{tabular}
\end{table}

\subsection{Workload and Configurations}
The workload is the motorBike tutorial of OpenFOAM v2112, a steady incompressible flow around a motorbike and its rider solved with simpleFoam. Its unstructured mesh and solver family resemble those of the urban air pollution model of our earlier study \cite{law2025prediction}. The case setup and the eight meshes, from 0.17 to 39.4 million cells, follow our study of this workload on AMD EPYC processors \cite{law2025evaluating}. OpenFOAM is built with AOCC 5.0 \cite{AOCC} for each microarchitecture and linked against Open MPI 4.1.5. Scotch partitions each mesh into one subdomain per rank, so the decomposition changes with the number of ranks. A rankfile binds the same number of ranks to every CCX. The five placements use one half, three quarters, all, one and a quarter and one and a half times as many ranks as a CCX has cores, or one quarter to one and a quarter on Turin-c. Placements above one also use the second hardware thread of some cores. Every mesh runs on one, two and four nodes, or on one and two nodes on Turin-128. This gives 15 or 10 configurations per processor and mesh and 560 configurations in total.

\subsection{Measurements}
The runtime $T$ is the time of 199 solver steps between the first and the last reported step, averaged over five repetitions. Only this average was kept, so the variation between repetitions is not available. The power $P$ is the package power of all sockets in the job, reported every second by the processor telemetry and read with AMD $\mu$Prof \cite{uProf51} during the solver phase. It does not include memory modules, network adapters or fans. Runs on the smallest meshes last only a few seconds, so their power rests on few samples. The energy of a run is $E = P\,T$ and its cost in node hours is $H = N\,T$ for $N$ nodes. Separate runs recorded hardware counters such as the L3 miss rate, together with the MPI traffic per core, following the profiling method of \cite{law2025evaluating}. These quantities help to explain the results in Section~\ref{sec:regimes}. They are not model inputs, because they are unknown before a run.

\section{Speed and Cost Trade-off}
\label{sec:formulation}

A decision concerns one processor and one mesh. Its options $o$ are the measured combinations of node count and placement, each with a runtime $T_o$ and a cost $C_o$, where $C$ stands for either the node hours $H$ or the energy $E$. Let $f$ denote the fastest option and $c$ the cheapest one. Two numbers summarise the trade-off of a decision,
\begin{equation}
\Delta C = \frac{C_f}{C_c} - 1, \qquad \Delta T = \frac{T_c}{T_f} - 1,
\label{eq:price}
\end{equation}
namely the extra cost of full speed and the extra time of the cheapest option. Both are positive when $f \neq c$, which is when a decision has a trade-off. A user may also accept a slowdown $s$ and ask for the cheapest option whose runtime stays within $(1+s)\,T_f$, or set a node budget $N_{\max}$ and ask for the fastest option on at most $N_{\max}$ nodes. The balanced option $b$ minimises $\max(T_o/T_f,\,C_o/C_c)$, the larger of its two relative losses. It formalises the choice made in our earlier study, where a setup slightly slower than the fastest one kept the cost close to the cheapest one \cite{law2025prediction}. For each objective, namely the shortest runtime, the fewest node hours, the least energy or the balanced option, the optimal configuration is the option that minimises it. Five processors and eight meshes give 40 decisions.

\section{Measured Trade-off}
\label{sec:measured}

\begin{figure}[t]
\centering
\includegraphics[width=\textwidth]{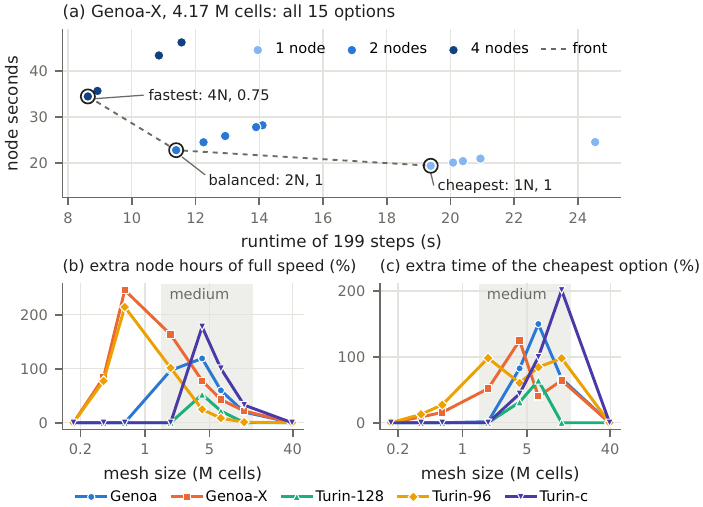}
\Description{Three panels. Panel a is a scatter plot of the cost in node seconds against the runtime for the 15 configurations of Genoa-X on the mesh of 4.17 million cells, coloured by node count. A dashed front links the fastest option (four nodes, three quarters of each CCX populated), the balanced option (two nodes, all cores) and the cheapest option (one node, all cores). Panels b and c plot the extra node hours of full speed and the extra time of the cheapest option against the mesh size on a logarithmic axis, one line per processor. Both quantities are zero for the smallest and the largest mesh and reach their highest values mostly for the medium meshes, which are shaded. In panel b Genoa-X and Turin-96 also show high values on small meshes.}
\caption{Example of the trade-off (a) and its two numbers for every processor and mesh (b, c).}
\label{fig:fronts}
\end{figure}

Figure~\ref{fig:fronts}a shows one decision, Genoa-X with the mesh of 4.17 million cells, with the cost in node seconds. Each of its 15 options appears as a point of runtime and cost. The dashed front connects the options that no other option outperforms in both runtime and cost. Its labels give the node count followed by the fraction of CCX cores that run a rank. Full speed on four nodes requires 78\% more node hours than the cheapest option on one node, whose runtime is 2.25 times longer. The balanced option on two nodes lies between them. Figures~\ref{fig:fronts}b and \ref{fig:fronts}c show the two numbers of Eq.~\eqref{eq:price} for every decision, summarised per mesh in Table~\ref{tab:regimes}. In about half of the decisions both numbers are zero, because one option is both the fastest and the cheapest in node hours.

\begin{table}[t]
\caption{Measured trade-off per mesh (medians over the five processors, node counts as ranges).}
\label{tab:regimes}
\centering
\setlength{\tabcolsep}{4pt}
\begin{tabular}{rcccccc}
\toprule
Cells & Fastest & Cheapest & \multicolumn{2}{c}{$\Delta C$ of full speed (\%)} & \multicolumn{2}{c}{$\Delta T$ of cheapest (\%)} \\
(M) & $N$ & $N$ (node h) & node hours & energy & node hours & energy \\
\midrule
0.17 & 1 & 1 & 0 & 0 & 0 & 0 \\
0.36 & 1--2 & 1 & 0 & 12 & 0 & 9 \\
0.60 & 1--4 & 1 & 0 & 13 & 0 & 9 \\
1.90 & 1--4 & 1 & 97 & 96 & 2 & 2 \\
4.17 & 2--4 & 1--2 & 78 & 76 & 60 & 60 \\
6.66 & 2--4 & 1--2 & 43 & 43 & 84 & 84 \\
11.9 & 2--4 & 1--2 & 20 & 18 & 67 & 69 \\
39.4 & 2--4 & 2--4 & 0 & 2 & 0 & 2 \\
\bottomrule
\end{tabular}
\end{table}

Within one to four nodes, the mesh size largely determines which of three regimes a decision belongs to. On three of the five processors small meshes are fastest and cheapest on one node, because additional nodes only add communication. Genoa-X and Turin-96 behave differently and gain speed from more nodes even on small meshes, at a substantial cost in node hours. Medium meshes from 1.9 to 11.9 million cells show a real trade-off on most processors. Full speed costs a median of 20 to 97\% more node hours than the cheapest option, which runs on one or two nodes and can take up to three times as long. On the mesh of 1.9 million cells the additional cost yields little reduction in runtime, since the cheapest option is only a median 2\% slower. For the largest mesh a single option on the largest node count is again both the fastest and the cheapest, because its speedup exceeds the number of nodes (Section~\ref{sec:regimes}). With more than four nodes a trade-off may appear for this mesh as well.

The size of the trade-off also depends on the processor. On Turin-128 and Turin-c the fastest option is usually the cheapest one as well, whereas full speed on Genoa-X costs a median 60\% more node hours. Accepting a longer runtime yields small savings, on average about a tenth of the node hours of full speed for a slowdown of 25\%. A budget of one node is far more restrictive and lengthens the runtime by a median 48\%. The balanced option removes most of the extra cost of full speed for a moderate loss of time. On the meshes of 4.2 and 6.7 million cells it saves about three quarters of the extra node hours while running about a third slower than full speed.

The placement matters independently of the node count. At a fixed node count the slowest placement takes a median 1.34 times as long as the fastest one. For the shortest runtime, filling every core of a CCX is best in only about a quarter of the cases. Half populated CCXs perform best on the small meshes, where each rank then has more L3 cache. Measured power also changes the conclusions of the earlier estimate based on TDP, which deviates from the measured energy by up to a third and moves the energy optimum in 15 of the 40 decisions.

\section{Hardware Explanation}
\label{sec:regimes}

\begin{figure}[t]
\centering
\includegraphics[width=\textwidth]{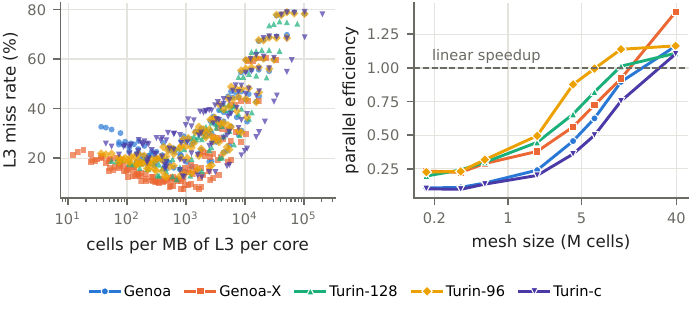}
\Description{Two panels. The left panel is a scatter plot of the L3 miss rate of all 560 runs against the number of cells per megabyte of L3 cache per core on a logarithmic axis, coloured by processor. The miss rate stays near 20 percent up to about a thousand cells per megabyte and rises to about 80 percent for the largest subdomains. The right panel plots the parallel efficiency at the largest node count against the mesh size, one line per processor. The efficiency rises from 0.1 to 0.2 for the smallest mesh to between 1.1 and 1.4 for the largest mesh and crosses the dashed line of linear speedup between about ten and forty million cells.}
\caption{L3 miss rate of all runs (left) and parallel efficiency at the largest node count (right).}
\label{fig:regimes}
\end{figure}

Figure~\ref{fig:regimes} relates these observations to the memory hierarchy. Its left panel shows that the L3 miss rate rises steeply with the number of cells per megabyte of L3 cache per core, from about 20\% for subdomains that fit into the cache to more than 70\% for the largest ones. This agrees with our earlier comparison of EPYC architectures \cite{law2025evaluating}. Adding nodes splits the mesh into smaller subdomains and moves a run to the left along this curve.

The right panel shows the parallel efficiency $S_N/N$ at the largest node count, which is four nodes or two on Turin-128. Here $S_N$ is the speedup of the fastest placement on $N$ nodes over the fastest placement on one node. For the smallest meshes the largest node count is even slower than a single node, because each rank holds only a few hundred cells and the halo exchange dominates. The efficiency grows with the mesh size and exceeds one for the largest mesh on every processor, reaching 1.42 on Genoa-X. For this mesh the L3 miss rate drops markedly with more nodes, as expected when the cache causes superlinear speedup \cite{ristov2016superlinear}. On most processors the miss rate on four nodes is still high, however, so the cache explains only part of the gain.

Over the 72 combinations of processor, mesh and a node count above one, the parallel efficiency correlates with the change in L3 miss rate relative to one node ($r=-0.61$). It correlates more strongly with the MPI traffic per cell ($r=-0.85$), which reflects the ratio of halo to volume of a subdomain. Since both quantities change with the number of cells per rank, the correlations are consistent with both mechanisms without separating them. For medium meshes the two mechanisms compete, since more nodes still reduce cache misses but also add halo traffic. Speed and cost therefore conflict in this range, where the user faces a genuine choice that a prediction can support.

\section{Prediction Method}
\label{sec:method}

\subsection{Proposed Procedure}
The method addresses a user with a new mesh of an already measured case on one of the processors in Table~\ref{tab:cpus}. The other meshes of the case have been measured on this processor, but the new one has not. For each objective the method first recommends the configuration that was optimal on the measured mesh closest to the new one in the logarithm of the cell count, on the same processor. This rule relies only on measured data and transfers the node count together with the placement.

Next, the user runs the configurations recommended for the shortest runtime and for the fewest node hours on the new mesh. The faster run provides $T_f$ and $H_f$ while the other provides $T_c$ and $H_c$. From these values Eq.~\eqref{eq:price} gives the estimated trade-off of the new mesh. If both recommendations coincide, a single run suffices. If the faster run is also the cheaper one, the method reports no trade-off. The procedure thus needs at most two runs of the new mesh.

\subsection{Reference Predictors}
Gradient boosted trees (XGBoost \cite{chen2016xgboost}) serve as a surrogate model. They predict the runtime and power of every option of the new mesh from descriptors known before the run, namely the hardware parameters of Table~\ref{tab:cpus}, the node count, the placement, the mesh size and derived ratios such as cells per core and cells per megabyte of L3 per core. One model per target is trained on all runs of the other meshes plus the $k$ runs of the new mesh on the processor in question. The measured values of these runs replace their predictions. The main surrogate uses a default configuration (500 trees of depth six, learning rate 0.05), which avoids any tuning on the withheld mesh. A configuration tuned by random search on the whole data set serves as a check. From the predicted costs the surrogate derives a recommendation and a trade-off with the same rules as for measured data. The earlier model \cite{law2025prediction} is reproduced for each processor. It fits CVOPTS as a polynomial of degree six in the logarithm of cells per rank, using the runs of the other meshes on two and four nodes. It is used in its original form, which scales the fit by the number of cell faces on the boundaries between subdomains that carry the MPI traffic, as well as in a variant with one polynomial per node count. Its energy estimate is $2\,N\,\mathrm{TDP}\,T$ for $N$ nodes with two sockets each. Without any run, the nearest measured mesh also provides a trade-off, with its runtime scaled by the ratio of cell counts. Finally, a user may select the best of $k$ random runs or apply one of two fixed rules, filling every core on one node or filling every core on the node count that gives about $3\times10^4$ cells per core.

\section{Example Predictions}
\label{sec:predictions}

Table~\ref{tab:pred} shows the output of the method for Genoa when each mesh is treated as new. Configurations are written as in Fig.~\ref{fig:fronts}a.

\begin{table}
\caption{Recommended configurations and estimated trade-off for Genoa as a new mesh.}
\label{tab:pred}
\centering
\setlength{\tabcolsep}{4pt}
\begin{tabular}{rcccccc}
\toprule
Cells & \multicolumn{3}{c}{Recommended configuration} & Runs & $\Delta C$ & $\Delta T$ \\
(M) & fastest & fewest node hours & balanced & & node hours (\%) & (\%) \\
\midrule
0.17 & 1N, 0.75 & 1N, 0.75 & 1N, 0.75 & 1 & 0 & 0 \\
0.36 & 1N, 0.75 & 1N, 0.75 & 1N, 0.75 & 1 & 0 & 0 \\
0.60 & 1N, 0.75 & 1N, 0.75 & 1N, 0.75 & 1 & 0 & 0 \\
1.90 & 4N, 0.5 & 1N, 1 & 2N, 0.75 & 2 & 0 & 0 \\
4.17 & 4N, 0.5 & 1N, 1 & 2N, 0.75 & 2 & 119 & 83 \\
6.66 & 4N, 0.5 & 1N, 1 & 2N, 0.75 & 2 & 60 & 150 \\
11.9 & 4N, 0.5 & 1N, 1 & 2N, 0.75 & 2 & 12 & 258 \\
39.4 & 4N, 0.5 & 2N, 1 & 4N, 0.5 & 2 & 0 & 0 \\
\bottomrule
\end{tabular}
\end{table}

For the three small meshes the nearest mesh recommends one node with three quarters of the cores for every objective, so a single run suffices and no trade-off is expected. For the larger meshes the recommendations for speed and cost differ, so two runs follow. The mesh of 4.17 million cells shows a clear trade-off. Full speed on four nodes costs 119\% more node hours there, while a single node takes 83\% longer. For 6.66 million cells the extra cost decreases to 60\%, while a single node requires 2.5 times the runtime. At 11.9 million cells full speed costs only 12\% more node hours and is therefore the preferable option. For the meshes of 1.9 and 39.4 million cells the faster run is also the cheaper one, so the method reports no trade-off.

\section{Assessment}
\label{sec:assessment}

\subsection{Protocol}
\label{sec:protocol}
Each mesh is withheld in turn and treated as new, separately for every processor. The column ``New runs'' in Tables~\ref{tab:optimal} and \ref{tab:calc} counts the runs of the withheld mesh, whereas predictors with zero new runs rely only on the measurements of the other meshes. Random runs for the surrogate and for the best of $k$ runs are drawn ten times, independently for every processor, mesh and draw. A recommended configuration counts as near optimal when its measured objective lies within 5\% of the optimum of the decision. For node hours the trade-off is scored by the absolute error of $\Delta C$ and $\Delta T$ in percentage points on the 21 decisions with a measured trade-off. The share of all 40 decisions in which the presence of a trade-off ($\Delta C > 5\%$) is classified correctly is reported as well. Differences between predictors are tested with an exact sign-flip test over meshes. Its smallest attainable $p$ value is 0.008 for eight meshes and 0.031 for the six meshes that contain a trade-off.

\subsection{Recommended Configurations}
Table~\ref{tab:optimal} gives the share of near optimal recommendations for the four objectives.

\begin{table}
\caption{Near optimal recommendations for a new mesh (\% of the 40 decisions).}
\label{tab:optimal}
\centering
\setlength{\tabcolsep}{4pt}
\begin{tabular}{lccccc}
\toprule
 & New & & Node & & \\
Predictor & runs & Runtime & hours & Energy & Balanced \\
\midrule
Nearest mesh & 0 & 75 & 78 & 85 & 75 \\
Nearest mesh, guided & 2 & 85 & 90 & 85 & 75 \\
Surrogate & 0 & 43 & 25 & 40 & 30 \\
Surrogate & 3 random & 57 & 49 & 60 & 45 \\
Surrogate & 2 guided & 53 & 48 & 53 & 45 \\
Surrogate, tuned & 3 random & 65 & 54 & 60 & 54 \\
Best of random runs & 3 random & 33 & 42 & 44 & 37 \\
Polynomial model & 0 & 13 & 20 & 8 & 8 \\
Polynomial per $N$ & 0 & 5 & 13 & 5 & 5 \\
All cores, one node & 0 & 13 & 50 & 50 & 30 \\
Cells per core rule & 0 & 20 & 50 & 58 & 30 \\
\bottomrule
\end{tabular}
\end{table}

Without any run, the configuration of the nearest measured mesh is already near optimal in 75 to 85\% of the decisions. The two guided runs raise this share to 85 to 90\% for runtime and node hours, because the better of the two measured options can be chosen. The surrogate remains well below both, whether it receives three random runs or the same two guided runs. Compared with the surrogate on three random runs, the method is near optimal 25 to 41 percentage points more often, which is significant for runtime, node hours and energy ($p=0.008$). Tuning the surrogate increases these shares by a few points without changing the ranking. The best of three random runs, the polynomial models and the fixed rules are near optimal for the shortest runtime in at most a third of the decisions. The fixed rules perform better for the cost objectives, since one node is the cheapest option for small meshes.

\subsection{Estimated Trade-off}
\begin{figure}
\centering
\includegraphics[width=\textwidth]{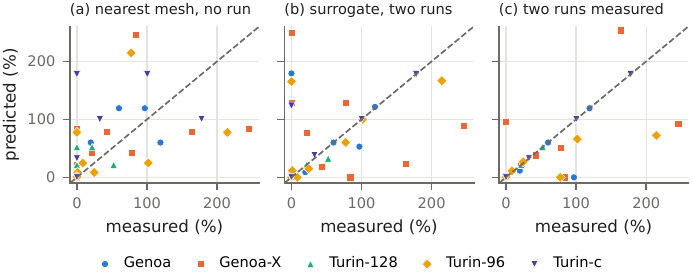}
\Description{Three scatter plots of the estimated against the measured extra node hours of full speed for the 40 decisions, coloured by processor, with the diagonal as reference. Panel a uses the nearest measured mesh without runs of the new mesh and scatters widely. Panel b uses the surrogate with two runs of the new mesh and lies closer to the diagonal. Panel c uses the two guided runs and lies on the diagonal for most decisions, with a few large deviations.}
\caption{Estimated and measured extra node hours of full speed for a new mesh.}
\label{fig:price}
\end{figure}

Table~\ref{tab:calc} lists the errors of the estimated trade-off, while Fig.~\ref{fig:price} compares the estimated and the measured $\Delta C$ for every decision. In a typical decision the two guided runs estimate the trade-off within a few points, with median errors of 3 and 5 points. They also detect correctly whether a trade-off exists in 90\% of the decisions. Their mean errors of 33 to 35 points arise from a few decisions in which the nearest mesh indicates the wrong pair, visible as the points off the diagonal in Fig.~\ref{fig:price}c. Without any run, the median errors of the nearest measured mesh are 38 to 59 points, since the size of the trade-off varies from one mesh to the next. Without runs of the new mesh the surrogate is no more accurate than the nearest measured mesh. With runs its median errors decrease but remain several times larger than those of the guided runs, even when it receives the same two runs. Compared with the surrogate on three random runs, the guided runs are significantly better for $\Delta T$ ($p=0.031$). For $\Delta C$ the medians differ clearly while the per-mesh means differ by only 5 points, so the difference is not significant. The median errors of the polynomial models are 31 to 78 points.

\begin{table}
\caption{Error of the estimated trade-off for a new mesh (node hours).}
\label{tab:calc}
\centering
\setlength{\tabcolsep}{4pt}
\begin{tabular}{lcccccc}
\toprule
 & New & \multicolumn{2}{c}{$\Delta C$ (pp)} & \multicolumn{2}{c}{$\Delta T$ (pp)} & Presence \\
Predictor & runs & median & mean & median & mean & correct (\%) \\
\midrule
Nearest mesh & 0 & 59 & 65 & 38 & 47 & 78 \\
Nearest mesh, guided & 2 & 3 & 35 & 5 & 33 & 90 \\
Surrogate & 0 & 52 & 70 & 56 & 67 & 58 \\
Surrogate & 3 random & 25 & 45 & 39 & 60 & 74 \\
Surrogate & 2 guided & 11 & 33 & 20 & 54 & 80 \\
Surrogate, tuned & 3 random & 22 & 41 & 31 & 56 & 75 \\
Polynomial model & 0 & 78 & 78 & 31 & 44 & 60 \\
Polynomial per $N$ & 0 & 53 & 62 & 50 & 57 & 73 \\
\bottomrule
\end{tabular}
\end{table}

\subsection{Example Check}
For the Genoa example in Table~\ref{tab:pred}, the recommended node count matches the measured optimum on 7 of the 8 meshes for the shortest runtime and on 6 for the fewest node hours and for the balanced option. The fastest and the cheapest recommendations are within 8\% of the optimum. The estimated trade-off is exact for six meshes. For the mesh of 1.9 million cells the measured fastest option on two nodes is only 1.5\% faster than the cheapest one, so reporting no trade-off has little consequence. For 11.9 million cells the measured cheapest option runs on two nodes while the recommended one uses a single node, so the extra time of the cheap choice is overestimated. Across the five processors the recommended node count is correct for 5 to 7 of the 8 meshes.

\subsection{Limits of the Surrogate}
On a new mesh without runs the surrogate predicts the runtime with a median error of 18 to 19\%, mostly because it has to extrapolate in mesh size for the smallest and the largest mesh. The best options of a decision, however, are often only a few percent apart. In 53\% of the decisions the second fastest option is within 5\% of the fastest one. For node hours this holds in 78\% of the decisions. Errors of almost a fifth can therefore change the order of such options. The nearest measured mesh avoids this problem, because neighbouring meshes usually share their optimal configuration. These near ties also limit all scores in this section, since averages over five repetitions cannot show whether such small differences exceed the variation between runs.

\section{Conclusions}
\label{sec:conclusions}

This paper revisited the performance and energy trade-off of memory-bound CFD on AMD EPYC clusters with measured power, five processors of two generations and the rank placement as a second decision. Within one to four nodes a real trade-off between speed and cost appears mainly for medium meshes, where full speed costs a median 20 to 97\% more node hours than the cheapest option. Small meshes are best run on one node on most processors, whereas superlinear speedup makes the largest node count both the fastest and the cheapest choice for the largest mesh. The placement alone changes the runtime by a median factor of 1.34 at a fixed node count. Filling every core of a CCX is rarely the fastest choice.

For a new mesh of a measured case the data support a simple procedure. Taking the configuration that was optimal on the nearest measured mesh already gives a near optimal choice in 75 to 85\% of the decisions. Running the two configurations it recommends for speed and for cost raises this share to 85 to 90\% for runtime and node hours and determines the trade-off with median errors of 3 to 5 percentage points. A surrogate model trained on the same data does not improve on this procedure, even with the same runs. The earlier polynomial model and common fixed rules perform clearly worse.

The study covers one CFD case, up to four nodes and the averages of five repetitions. Future work will measure the variation between repetitions, extend the node counts and test whether the configurations transfer between different CFD cases.

\begin{credits}
\subsubsection{\ackname}
Funded by the European Union. This work has received funding from the European High Performance Computing Joint Undertaking and Poland, Germany, Spain, Hungary, France and Greece under grant agreement number: 101093457. This publication expresses the opinions of the authors and not necessarily those of the EuroHPC JU and Associated Countries which are not responsible for any use of the information contained in this publication. The authors thank the AMD company for granting access to the HPC platforms.

\subsubsection{\discintname}
The authors have no competing interests to declare that are relevant to the content of this article.
\end{credits}

\end{document}